\documentclass[aps,prl,twocolumn,showpacs,superscriptaddress,floatfix,10pt]{revtex4-1}
\usepackage{framed}
\usepackage{graphicx}
\usepackage{amsmath}
\usepackage{epsfig}
\usepackage{helvet}
\usepackage{amssymb}
\usepackage{framed}
\usepackage[plainpages=false,colorlinks=true,linkcolor=blue, citecolor=blue, urlcolor=blue]{hyperref}

\newcommand{\ket}[1]{\mbox{$| #1 \rangle$}}

\newcommand{\frw}[1]{$\overset{\lower0.5em\hbox{$\smash{\scriptscriptstyle\smile}$}} #1$}

\def\D{{\mathcal D}}

\def\E{{\mathcal E}}

\def\L{{\mathcal L}}

\def\C{{\mathcal C}}
\def\R{{\mathcal R}}

\begin{document}

\title{Hyper-invariant tensor networks and holography}

\author{Glen Evenbly}
\affiliation{D\'epartement de Physique and Institut Quantique, Universit\'e de Sherbrooke, Qu\'ebec, Canada
}
\email{glen.evenbly@usherbrooke.ca}
\date{\today}

\begin{abstract}
We propose a new class of tensor network state as a model for the AdS/CFT correspondence and holography. This class is demonstrated to retain key features of the multi-scale entanglement renormalization ansatz (MERA), in that they describe quantum states with algebraic correlation functions, have free variational parameters, and are efficiently contractible. Yet, unlike MERA, they are built according to a uniform tiling of hyperbolic space, without inherent directionality or preferred locations in the holographic bulk, and thus circumvent key arguments made against the MERA as a model for AdS/CFT. Novel holographic features of this tensor network class are examined, such as an equivalence between the causal cones $\C(\R)$ and the entanglement wedges $\E(\R)$ of connected boundary regions $\R$.
\end{abstract}

\maketitle

\textbf{Introduction.---} 
Tensor network methods \cite{Cirac09,Evenbly11a} have proven remarkably useful for investigating quantum many-body systems, both advancing their theoretical understanding and providing powerful tools for their numeric simulation. Introduced by Vidal, the multi-scale entanglement renormalization ansatz (MERA) \cite{Vidal08}, which describes quantum states on a $D$-dimensional lattice as a tensor network in ($D$+1)-dimensions, is known to be particularly well-suited for representing ground states of critical systems \cite{Gio08, Pfeifer09, Evenbly10d, Evenbly11b, Bridgeman15}, such as lattice versions of conformal field theories (CFTs) \cite{CFT1, CFT2}. Importantly, by imposing invariance along the emergent dimension of the network, which can be regarded as a renormalization scale \cite{ER}, the scale-invariance of critical systems can be captured by MERA.

More recently tensor networks have also emerged in the study of the AdS/CFT correspondence \cite{Maldacena98, Witten98, Ryu06}, and of holography in general. The AdS/CFT correspondence, a duality between quantum gravity on a ($D$+1)-dimensional AdS space and a $D$-dimensional CFT defined on its boundary, has offered new insights into both quantum gravity and strongly-coupled quantum field theories. Observing that MERA have hyperbolic geometries similar to a spatial slice of AdS, in conjunction with the numeric success of MERA in encoding ground states of CFTs, it was argued by Swingle that MERA may realize key aspects of 
holography \cite{Swingle12, Swingle12b}. This observation has since generated a great deal of interest \cite{Taka12, Qi13, Beny13, Evenbly15, Czech15, Bao15, Miyaji16, Czech16, Czech16b, Chua16, Hayden16, Bhatta16, Singh17}, both within the AdS/CFT and tensor network communities.

Although the proposal that tensor networks capture some aspects of holography has been undoubtedly useful, for instance, in stimulating the development of new tensor network methods,
several works have argued against MERA as a direct realization of the AdS/CFT duality \cite{Beny13, Czech15, Bao15}. Many of the significant criticisms ultimately stem from the problem that, when viewed as a tiling of hyperbolic space, MERA have preferred directions resulting from their use of unitary and isometric tensors, in contrast to the uniform AdS bulk. This concern was a motivating factor for the introduction of holographic codes as models for the AdS/CFT correspondence \cite{Pasta15}, which use so-called \emph{perfect tensors} to construct a family of tensor networks that are uniform in the holographic bulk. However, it is known that holographic codes cannot be related to ground states of critical systems, as they do not produce compatible correlation functions or entanglement spectra.

\begin{figure}[!t!b]
\begin{center}
\includegraphics[width=8.6cm]{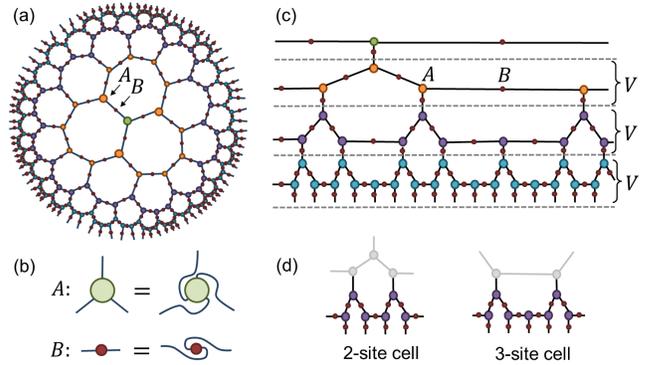}
\caption{(a) A hyper-invariant tensor network based on a $\{7,3\}$ tiling of the hyperbolic disk, where a 3-index tensor $A$ is placed on each node of the tiling, with a matrix $B$ then placed on each edge adjoining two nodes. (b) Tensors $A$ are constrained to be rotationally invariant, while matrices $B$ are constrained to be symmetric. (c) The network can be unwrapped into concentric layers $V$ about any chosen center, with each layer as a connected string of alternating $A$ and $B$ tensors. (d) All layers (with the exception of that immediately following the top) are formed from a combination of 2-site and 3-site unit cells. }
\label{fig:Tiling37}
\end{center}
\end{figure}

Thus, it remains an interesting open question: does there exist a class of tensor network that is both uniform in the holographic bulk and can also produce correlations/entanglement compatible with critical ground states? The purpose of the present manuscript is to answer this question in the affirmative, with the introduction of \emph{hyper-invariant} tensor networks, which capture the desirable aspects of both the MERA and holographic codes as models for holography. Specifically, they (i) are built from a uniform tiling of hyperbolic space (i.e. with no inherent directionality or preferred locations), (ii) are efficiently contractible, (iii) have free variational parameters, and (iv) encode quantum states with algebraic decay of two-point correlation functions. 

\textbf{Construction.---} We build a hyper-invariant tensor network according to a \emph{hyperbolic tessellation} or uniform tiling of the hyperbolic disk. For simplicity, we shall focus on a $\{7,3\}$ tessellation, i.e. a tiling with 7-edged plaquettes and 3-edged nodes as depicted in Fig. \ref{fig:Tiling37}(a) (although an alternative network based on a $\{5,4\}$ tiling is considered in Sect. C of the supplemental material). In this construction a 3-index tensor $A$ is placed on each node of the tiling and a matrix $B$ is placed on each edge adjoining two nodes. In order to be compatible with bulk uniformity, we constrain the tensor $A$ to be invariant with respect to a cyclic index permutation and constrain matrix $B$ to be symmetric, see Fig. \ref{fig:Tiling37}(b).

It is useful to organise the network into concentric layers $V$ around the $A$ tensor at a chosen bulk point $T$, with each layer a connected string of alternating $A$ and $B$ tensors as depicted in Fig. \ref{fig:Tiling37}(c). One can then regard a layer $V$ as defining a renormalization group (RG) transformation from a $1D$ lattice $\L_z$ to a coarser lattice $\L_{z+1}$, where $z$ is a label over scale that increases moving towards $T$. In contrast to the MERA, each layer $V$ here does not consist of translations of a single unit cell but instead is composed of a combination of 2-site and 3-site unit cells, see Fig. \ref{fig:Tiling37}(d). Notice that each 2-site cell sits underneath an arrangement of three tensors from the preceding layer, whilst each 3-site cell sits underneath a pair of tensors from the preceding layer (consistent with the property that all plaquettes are 7-edged). It follows that the pattern of cells is fractal in nature, such that there is no finite repeating pattern of cells even in the limit of a layer infinitely far from the center. Consequentially, the ratio $r$ of 3-site to 2-site unit cells and the scale factor $s$ (i.e. the ratio of sites in $\L_z$ to that in the coarser lattice $\L_{z+1}$) are both irrational in the thermodynamic limit,
\begin{equation}
r = (1 + \sqrt 5)/2 \approx 1.618, \; \; \; s = 1+r \approx 2.618. \label{eq:scalefactor}
\end{equation}

\begin{figure}[!t!b]
\begin{center}
\includegraphics[width=8.6cm]{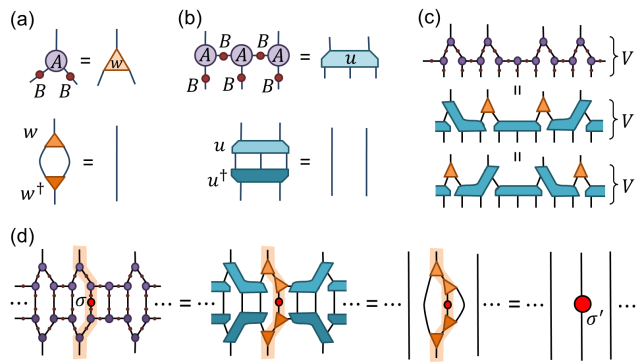}
\caption{(a) The product of an $A$ and two $B$ tensors is constrained to form an isometry $w$ which, by definition, annihilates to identity with its conjugate $w^\dag$. (b) Tensor $u$, which is formed from three $A$ tensors together with five $B$ tensors, is constrained to be an isometric mapping from 3-to-2 indices. (c) The tensors in each layer $V$ can be grouped as a product of $w$ and $u$ isometries in many different ways. (d) Under the action of layer $V$, a one-site local operator $\sigma$ is mapped to a coarse-grained operator, $\sigma' \equiv {V^\dag }\sigma V$, which remains local due to the cancellation of tensors in $V$ with their conjugates in $V^\dag$ (where tensors have been grouped into isometries $w$ and $u$ in such a way as to minimize the support of $\sigma'$).}
\label{fig:Constraint37}
\end{center}
\end{figure}

\textbf{Multi-tensor constraints.---} Fundamental in MERA is the use of isometric and unitary tensors which ensure, when interpreting each layer as a transformation from an initial lattice $\L_z$ to a coarser lattice $\L_{z+1}$, that local operators are mapped to local operators \cite{ER}. For instance, in the case of a $1D$ binary MERA, any local operator $\sigma$ supported on $L\le 3$ neighboring sites of $\L_z$ is mapped to a local operator $\sigma'$ on supported on $L \le 3$ neighboring sites of the coarser lattice $\L_{z+1}$. The property of preserving locality as a coarse-graining transformation, or equivalently, that MERA have \emph{bounded causal width} \cite{Vidal08}, is key not only to their efficient contraction for local expectation values and correlation functions, but also to their ability to reproduce expected features of CFTs such as scaling operators and their fusion coefficients \cite{Gio08, Pfeifer09, Evenbly10d}. Holographic codes \cite{Pasta15}, on the other hand, make use of \emph{perfect tensors}, which are isometric across all possible partitions of indices, to achieve bulk uniformity while also preserving locality. However, when viewed as a coarse-graining transformation between initial $\L_z$ and coarser $\L_{z+1}$ lattices, the use of perfect tensors results in regions on $\L_z$ for which any local operator $\sigma$ supported on the region is coarse-grained to the trivial (i.e. identity) operator on $\L_{z+1}$, which implies the existence of trivial connected correlation functions in the holographic codes.

Thus, in order to be compatible with both bulk uniformity and the preservation of locality, yet still allow for non-trivial correlation functions, a different type of tensor constraint is needed, which we now describe. Instead of attempting to constrain individual tensors within the network, as done in both MERA and the holographic codes, here we propose the use of \emph{multi-tensors constraints}, which constrain how certain products of tensors behave in conjunction with one another. Specifically, for the $\{7,3\}$ hyper-invariant network, the product of an $A$ and two $B$ tensors is constrained to act as a 2-to-1 isometry $w$, while a product of three $A$ and five $B$ tensors is constrained to act as a 3-to-2 isometry $u$, as depicted in Fig. \ref{fig:Constraint37}(a-b). We defer to Sect. B of the supplemental material for a demonstration of the existence of tensors $A$ and $B$ 
that satisfy these constraints, and discussion on how solutions can be realised in general.

\begin{figure}[!t!b]
\begin{center}
\includegraphics[width=8.6cm]{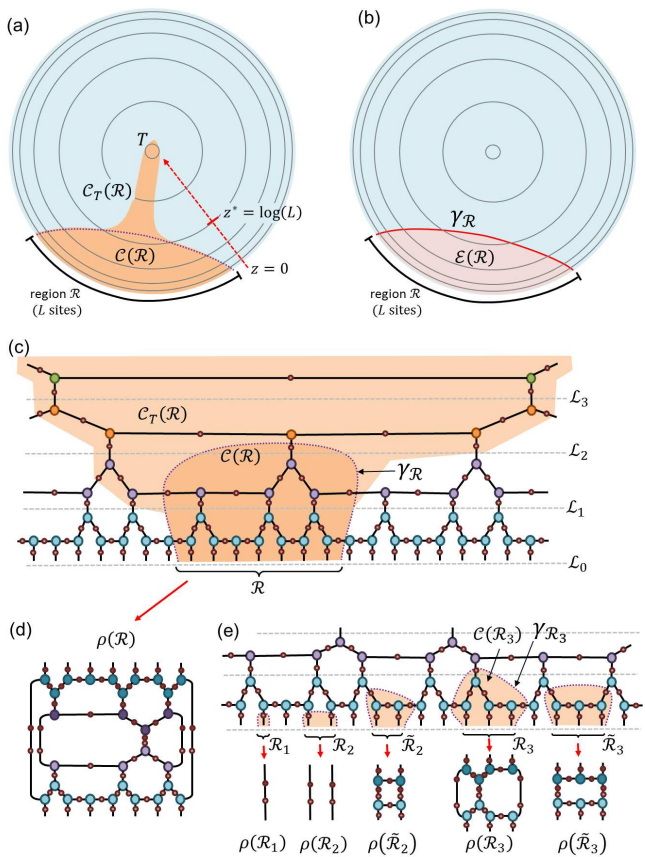}
\caption{(a) A schematic representation of a the hyper-invariant network, which has been organised into layers about a chosen bulk point $T$. The apparent causal cone $\C_T(\R)$ of a boundary region $\R$ of $L$ sites shaded, where the true causal cone $\C(\R)$ is a subset of $\C_T(\R)$. (b) The entanglement wedge $\E(\R)$ is the bulk region bounded by the minimal surface $\gamma_\R$ and $\R$. (c-d) Part of a $\{7,3\}$ hyper-invariant network that has been organised into layers, with the apparent causal cone $\C_T(\R)$ of a boundary region $\R$ shaded. In the evaluation of the reduced density matrix $\rho(\R)$ many tensors in $\C_T(\R)$ cancel, such that $\rho(\R)$ only depends on tensors within a subset $\C(\R) \subset \C_T(\R)$, where it is observed that $\C(\R)$ is exactly coincident with the entanglement wedge $\E(\R)$. (e) Depictions of the causal cones $\C(\R)$ and the corresponding reduced density matrices $\rho(\R)$ for one, two and three site regions $\R$, where the causal cones also correspond to entanglement wedges $\E(\R)$.}
\label{fig:Holographic}
\end{center}
\end{figure}

We now examine the implications of these constraints on the hyper-invariant network, which we organise into layers $V$ about a chosen bulk point $T$ as depicted in Fig. \ref{fig:Tiling37}(c). We regard each layer $V$ as a mapping between an initial $\L_z$ and coarser lattice $\L_{z+1}$, where a local operator $\sigma$ on $\L_z$ is mapped to a coarser operator $\sigma'$ on $\L_{z+1}$ as
\begin{equation}
\sigma' \equiv {V^\dag }\sigma V. \label{eq:1}
\end{equation}
Notice that the constraints allow the tensors in a layer $V$ to be grouped into a product of $w$ and $u$ isometries in many different ways, see Fig. \ref{fig:Constraint37}(c). Under each grouping, many of the isometries $w$ and $u$ in a layer $V$ will cancel with their conjugates in $V^\dag$ in Eq. \ref{eq:1}. However, since no property of $\sigma'$ can depend on which grouping into isometries is imagined, it follows that the non-trivial part of $\sigma'$ should be understood from the grouping that results in the \emph{minimal} support, as depicted in Fig. \ref{fig:Constraint37}(d). Through examination of all possible ways in which local operators can be coarse-grained, see Sect. A of the supplemental material, it is seen that any local operator supported on $L\le 2$ sites of lattice $\L_z$ is mapped to a local operator $\sigma'$ supported on $L\le 2$ sites of the coarser lattice $\L_{z+1}$. Notice also that there does not exist regions on $\L_z$ for which operators are mapped to null regions on $\L_{z+1}$ (i.e. regions on $\L_z$ where \emph{any} operator on the region is mapped to the trivial identity operator). Thus the multi-tensor constraints achieve the desired goal of preservation of locality, hence also efficient contractibility of the networks, while not restricting to trivial correlation functions.

\textbf{Causal properties.---} As a direct consequence of bulk uniformity, the causal cones of hyper-invariant networks differ substainially from those of MERA. Let us assume that we have a finite hyper-invariant network, which describes a quantum state $\ket{\psi}$ on the lattice $\L$ associated to the boundary indices. For a region $\R \in \L$ the causal cone $\C(\R)$ is defined as the set of tensors in the bulk that can affect the density matrix $\rho(\R) = \textrm{tr}_{\bar{\R}}\left( \left| \psi  \right\rangle \left\langle \psi  \right| \right)$, where $\bar{\R}$ is the lattice compliment of $\R$. Additionally, following Ref. \cite{Pasta15}, we define the entanglement wedge $\E(\R)$ as the set of bulk tensors bounded by $\R$ and $\gamma_R$, where $\gamma_R$ is the minimal surface whose boundary matches the boundary of $\R$ \cite{geodesic}. The following relation between $\C(\R)$ and $\E(\R)$ is then observed:
\newline

\noindent \textit{\textbf{Holographic causality:} For a continuous boundary region $\R$ of a hyper-invariant network, the causal cone $\C(\R)$ is approximately coincident\textnormal{\cite{approx}} with the entanglement wedge $\E(\R)$.}
\newline

\noindent In order to understand this relation, it is first useful to introduce, given a hyper-invariant network that has been organised into layers $V$ about a bulk point $T$, the notion of an \emph{apparent} causal cone $\C_T(\R)$. Here $\C_T(\R)$ is defined as the minimal causal cone of a boundary region $\R$ that can be achieved from a layer-by-layer grouping of tensors into isometries $w$ and $u$, as depicted in Fig. \ref{fig:Constraint37}(c). The apparent causal cones $\C_T(\R)$ in hyper-invariant networks are then seen to take the same characteristic forms as causal cones in MERA \cite{Vidal08,Evenbly14}, see Fig. \ref{fig:Holographic}(a). Specifically, given a boundary region $\R$ of $L$ sites, assumed for simplicity to be at the edge of a complete layer $V$, the width of the cone diminishes by the scale factor ($s$ as defined in Eq. \ref{eq:scalefactor}) for each of the first $z^*\approx \log_s(L)$ layers into the bulk (the \emph{shrinking regime}), then remains at a small finite width for the remaining layers (the \emph{steady regime}). Thus, similar to observations about the causal cones in MERA \cite{Evenbly14}, here it is also observed that the shrinking regime of the apparent causal cone $\C_T(\R)$ is approximately coincident to the entanglement wedge $\E(\R)$. However, in the hyper-invariant network, tensors within the steady regime of $\C_T(\R)$ cancel in the evaluation of the density matrix $\rho(\R)$, such that the true causal cone $\C(\R)$ contains only the shrinking regime of $\C_T(\R)$. This is demonstrated in Sect. A of the supplemental material, where the dominant eigenoperators of the descending superoperators, which are used to evaluate the steady regime causal cones, are shown to evaluate trivially. Thus the property of holographic causality results, with $\C(\R) \approx \E(\R)$, as also demonstrated in the examples of Fig. \ref{fig:Holographic}.

Alternatively, holographic causality can also be understood as resulting from the freedom to organise the network into isometric layers $V$ about \emph{any} centre $T$. As depicted in Fig. \ref{fig:Center}, see also Sect. D of the supplemental material, by choosing $T$ at the apex of the minimal surface $\gamma_{\R}$ the apparent causal cone is minimized such that $\C_T(\R) = \C(\R)$, whereupon it is also seen that $\C(\R) \approx \E(\R)$. Finally we remark that holographic causality is not predicated on bulk uniformity; the causal properties remain the same even if tensors $A$ and $B$ are allowed to differ with location and direction in the bulk, so long as the multi-tensor constraints are still satisfied in all instances. 

\begin{figure}[!t!b]
\begin{center}
\includegraphics[width=8.0cm]{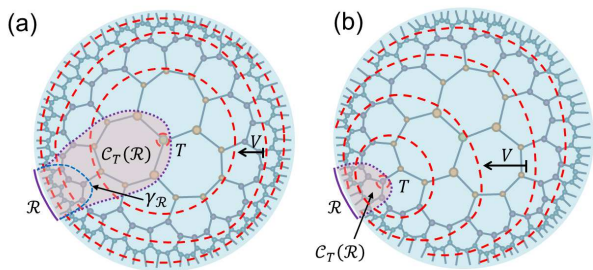}
\caption{(a) A hyper-invariant network organised into layers $V$ about bulk point $T$, with the apparent causal cone $\C_T(\R)$ and minimal surface $\gamma_{\R}$ of boundary region $\R$ shown. (b) Choosing point $T$ at the apex of $\gamma_{\R}$ gives $\C_T(\R) = \E(\R)$.}
\label{fig:Center}
\end{center}
\end{figure}

\textbf{Discussion.---} A noteworthy implication of holographic causality is that the volume of causal cones $\left\| \C(\R) \right\|$, measured by the number of tensors they contain, remain finite, with $\left\| \C(\R) \right\| \propto L$, for a continuous boundary region $\R$ of $L$ sites, even in the thermodynamic limit (i.e. of the network with infinite layers). Thus any local reduced density matrix $\rho(\R)$ from a hyper-invariant network has a closed-form expression as a finite network of $A$ and $B$ tensors (with their conjugates), as depicted in the examples of Fig. \ref{fig:Holographic}. This is in contrast to scale-invariant MERA whose causal cones include tensors at all scales such that $\left\| \C(\R) \right\| \rightarrow \infty$ for any boundary region $\R$, and no equivalent closed-form expressions for $\rho(\R)$ exist. Notice this further implies that, for any instance of a hyper-invariant network, the reduced density matrix $\rho(\R)$ for a boundary region $\R$ length of $L$ will evaluate to one of a finite number of unique $L$-site density matrices (i.e. for any choice of $\R$). For example, the reduced density matrices $\rho(\R_1)$ are found to be identical for any choice of $1$-site boundary region $\R_1$, see Fig. \ref{fig:Holographic}(e), while in MERA the $1$-site reduced density matrices would differ for all regions $\R_1$ as they each have unique causal cones through the bulk. This observation could suggest that hyper-invariant networks (though not, by default, translation invariant when viewed as quantum states on a $1D$ boundary lattice) are better suited for capturing translation invariance in quantum states than MERA.

In Sect. C of the supplemental material it is demonstrated that the hyper-invariant networks satisfy the key criteria necessary of variational tensor network ansatz. Namely they can be parameterized by a set of continuous parameters $\{\theta_1, \theta_2, \ldots, \theta_n \}$, where (i) the number of $n$ of free parameters increases with the bond dimension of the network, and (ii) the entanglement spectra and correlation functions depend non-trivially on these parameters. However, more work is required in order to understand what class of quantum system they may be suitable for as a ground state ansatz, and whether this includes critical lattice models. One could try to address this question numerically by building a variational algorithm to optimize hyper-invariant networks for ground states of lattice models. This would first require overcoming several obstacles, such as understanding (i) the best way parameterize solutions to multi-tensor constraints, (ii) how to incorporate transitional layers to act as a buffer between the physical lattice and the holographic bulk (as done in scale-invariant MERA algorithms \cite{Pfeifer09, Evenbly11b, Evenbly09}), and (iii) an approach to optimising the free parameters. However, if viable, such an algorithm could offer significant computational gains arising from the exploitation of bulk uniformity and from the use of the multi-tensor constraints, which simplify many of the required network contractions.

Similar to other tensor network proposals for holography \cite{Vidal08,Qi13,Singh17,Pasta15}, one could also add additional free indices to the bulk tensors of hyper-invariant tensor networks, such that they then realize a mapping from a $1D$ boundary lattice to a bulk (hyperbolic) lattice. This remains an interesting avenue for future work. Finally we note that some (perfect tensor) holographic states \cite{Pasta15} may be regarded as specific instances of hyper-invariant networks; for example a 4-index perfect tensor $A$ is also a valid solution to the multi-tensor constraints of the $\{5,4\}$ hyper-invariant network seen in Fig. \ref{fig:Constraint45} of Sect. C of the supplemental material. Thus, hyper-invariant networks (with the addition of free bulk indices, as discussed above) may also be useful in the construction of new holographic quantum error correcting codes beyond the paradigm of perfect tensors.

The author thanks Guifre Vidal, Fernando Pastawski and David Poulin for useful comments and discussions. 

\newpage

\textbf{$~~~~~~~~~$SUPPLEMENTAL MATERIAL}

\section{Section A: Descending superoperators and reduced density matrices}
\begin{figure}[!t]
\begin{center}
\includegraphics[width=8.6cm]{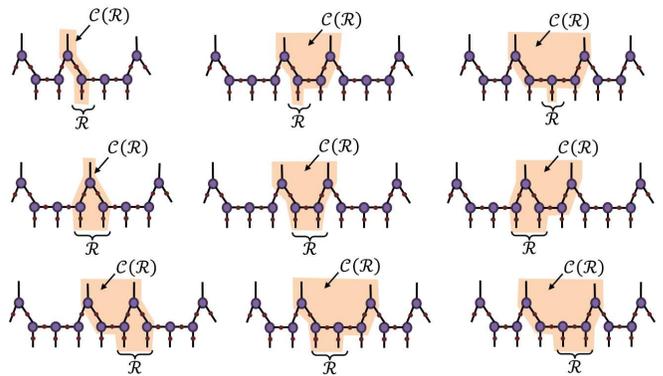}
\caption{Depictions of the causal cones $\C(\R)$ of 1-site and 2-site regions $\R$ through one layer of the $\{7,3 \}$ hyper-invariant network.}
\label{fig:CausalCones}
\end{center}
\end{figure}

In this section we further detail the evaluation of local reduced density matrices from the $\{7,3 \}$ hyper-invariant tensor network. We begin by investigating the causal properties of a layer $V$ of the network, which we interpret as defining a coarse-graining transformation between an initial lattice $\L_z$ and a coarser lattice $\L_{z+1}$, that are imposed through the multi-tensor constraints of Fig. \ref{fig:Constraint37}. As discussed in the main text, the causal cone $\C(\R)$ associated to region $\R \in \L_z$ through the layer $V$ contains the set of tensors that do not cancel in coarse-graining of operators $\sigma_{\R}$ supported on $\R$, 
\begin{equation}
{\sigma'} \equiv {V^\dag } \sigma V,
\end{equation}
where the tensors in $V$ are grouped into isometries $w$ and $u$ as to minimize the support of the coarse-grained operator ${\sigma'}$ on $\L_{z+1}$. Following this definition, the set of possible causal cones $\C(\R)$ of 1-site and 2-site regions $\R$ through a single layer $V$ are depicted in Fig. \ref{fig:CausalCones}, which are seen to remain as 1-site or 2-site regions.


Assume that we have a hyper-invariant network composed of some number of complete layers $V$ about a chosen bulk point $T$, with lattice $\L_0$ as the boundary lattice and $\L_z$ the lattice formed after $z$ layers of the network. We now discuss the evaluation a local reduced density matrix $\rho(\R)$ with $\R$ either a 1-site or 2-site region on $\L_0$. As introduced in the main text, it is useful here to define the \emph{apparent} causal cone $\C_T(\R)$ as the causal cone one sees through the network when applying the layer-by-layer causal analysis as discussed above. The local reduced density matrices from a hyper-invariant network can then be evaluated using the same formalism as that employed in MERA \cite{Evenbly09}; namely through the use of \emph{descending superoperators} $\D$. Here a reduced density matrix $\rho^{[z-1]}$ on lattice $\L_{z-1}$ can be obtained through application of the appropriate descending superoperator $\D$ on a density matrix $\rho^{[z]}$ on lattice $\L_{z}$. 
Thus, through repeated use of descending superoperators $\D$ the density matrix can be descended through the apparent causal cone $\C_T(\R)$, until the desired density matrix $\rho(\R)$ on the boundary lattice $\L_0$ is reached.

\begin{figure}[!t!b]
\begin{center}
\includegraphics[width=8.6cm]{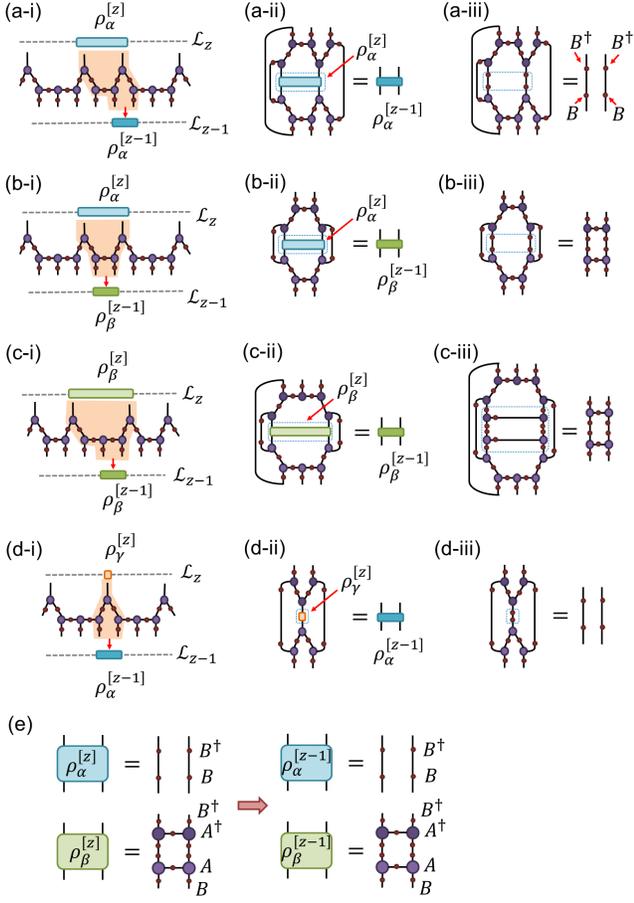}
\caption{(a-i) A two site density matrix $\rho_\alpha^{[z-1]}$ on lattice $\L_{z-1}$ can be obtained by descending the appropriate density matrix $\rho_\alpha^{[z]}$ on lattice $\L_{z}$ through the causal cone, (a-ii) implemented by a descending superoperator. (a-iii) The density matrix $\rho_\alpha = (B^\dag B) \otimes (B^\dag B)$ is seen to be a fixed point of the descending superoperator. (b-d) Other descending superoperators and their fixed points. (e) If the two site density matrices on lattice $\L_z$ are of the form depicted, with $\rho_\alpha$ and $\rho_\beta$ located under a configuration of three and two tensors from the preceding network layer respectively, then the two site density matrices on $\L_{z-1}$ take the same form.}
\label{fig:DescendingSupers}
\end{center}
\end{figure}

The various 2-site descending superoperators of the $\{7,3 \}$ hyper-invariant tensor network are depicted in Fig. \ref{fig:DescendingSupers}, where a significant difference from the descending superoperators of MERA can be observed. While the dominant eigenoperators of MERA \emph{ascending} superoperators are always known, as they correspond to identity operators, the dominant eigenoperators of MERA \emph{descending} superoperators are typically not known and can only be evaluated numerically. In contrast, through proper use of the multi-tensor constraints of Fig. \ref{fig:Constraint37}(a-b), the dominant eigenoperators of the descending superoperators from the hyper-invariant network \emph{can} be evaluated analytically. This is shown Fig. \ref{fig:DescendingSupers} where two different 2-site fixed point density matrices, $\rho_\alpha^{[z]}$ and $\rho_\beta^{[z]}$, are obtained depending on their position on $\L_z$ relative to the preceding layer $V$. Here the $\rho_\alpha^{[z]}$ sits below an arrangement of three $A$ tensors from the network above, while the $\rho_\beta^{[z]}$ sits below a pair of $A$ tensors. The explicit forms of $\rho_\alpha^{[z]}$ and $\rho_\beta^{[z]}$ are shown Fig. \ref{fig:Constraint37}(e). Important to note is they only depend on the $A$ and $B$ tensors from the layer $V$ of the network that immediately precedes lattice $\L_z$. In other words, all of the tensors in the apparent causal cone of a 2-site region $\R$ in $\L_z$, with the exception of those within the layer immediately preceding, annihilate with their conjugates in the evaluation of $\rho^{[z]}(\R)$. This observation, which implies that tensors in the \emph{steady} (i.e. fixed width) regime of an apparent causal cone $\C_T(\R)$ cancel in the evaluation of the local reduced density matrix, was central to the explanation of the holographic causality property of hyper-invariant tensor networks discussed in the main text.

\section{Section B: Solutions to the multi-tensor constraints}
In this section we provide an example of a family of tensors $A$ and $B$ that simultaneously satisfy the symmetry constraints of Fig. \ref{fig:Tiling37}(b) and the multi-tensor isometry constraints of Fig. \ref{fig:Constraint37}(a-b). Note that the example provided here is but one of many possible ways of parameterizing tensors which satisfy the constraints; it could be that other parameterizations are better suited for practical purposes (i.e. if the hyper-invariant tensor network is to be used as an ansatz for ground states). The purpose here is to demonstrate the existence of solutions to multi-tensor constraints for $A$ and $B$ that satisfy the following criteria:
\begin{enumerate}
\item they are parametrized by a set of continuous parameters $\{\theta_1, \theta_2, \ldots, \theta_n \}$,
\item the number $n$ of free parameters increases as the bond dimension of the network is increased,
\item they generate networks with non-trivial ($\theta$-dependent) entanglement spectra and correlation functions.
\end{enumerate}
These three properties are necessary for the hyper-invariant network to be considered as a potential variational ansatz for ground states of quantum systems.

\begin{figure}[!t!b]
\begin{center}
\includegraphics[width=8.6cm]{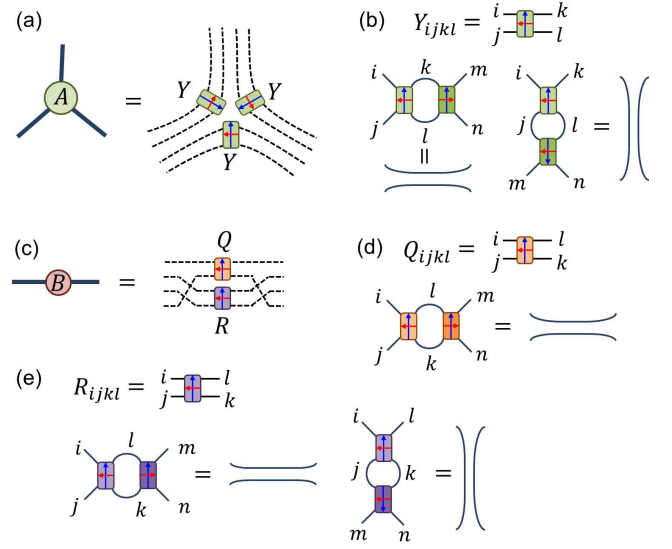}
\caption{A family of tensors $A$ and $B$ that are both compatible with bulk uniformity and satisfy the multi-tensor constraints of Fig. \ref{fig:Constraint37}(a-b) is constructed. (a) The indicies of tensors $A$ are decomposed as a product of four finer indices, such that $A$ is then given as the product of three tensors $Y$ (here the arrows denote the relative orientation of $Y$ tensors). (b) Tensors $Y$ are constrained to be \emph{doubly unitary}, i.e. simultaneously unitary across both horizontal and vertical partitions. (c)  The indicies of tensors $B$ are decomposed as a product of four finer indices, such that $B$ is then given as the product of tensors $Q$ and $R$. (d) Tensors $Q$ are constrained to be unitary across a single partition. (e) Tensors $R$ are constrained to be doubly unitary.}\label{fig:Parameterization37}
\end{center}
\end{figure}

\begin{figure}[!t!b]
\begin{center}
\includegraphics[width=8.6cm]{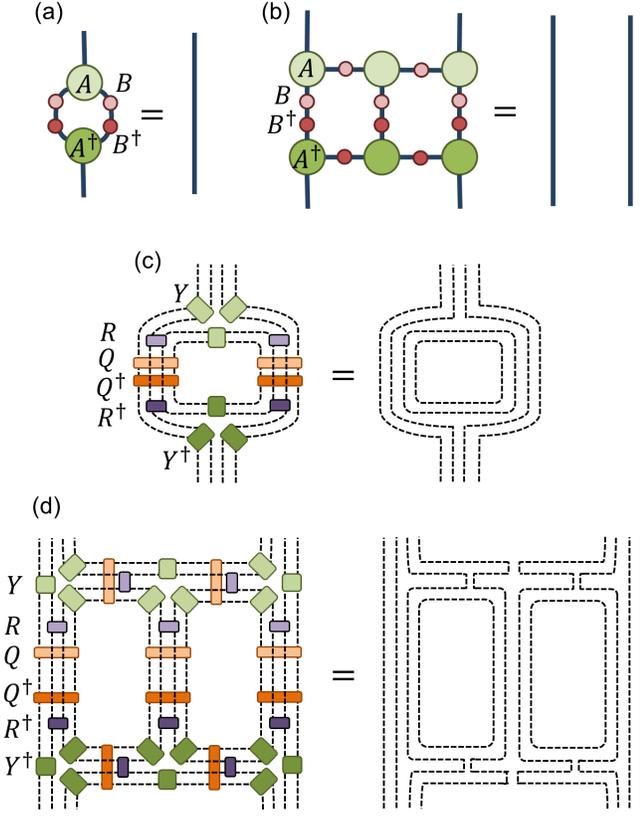}
\caption{(a-b) A depiction of the multi-tensor constraints imposed on the $\{7,3\}$ hyper-invariant network, see also Fig. \ref{fig:Constraint37}(a-b). (c-d) Tensors $A$ and $B$ that have been built with the internal structure described in Fig. \ref{fig:Parameterization37} are seen to exactly satisfy the multi-tensor constraints up to (unimportant) multiplicative constants.}
\label{fig:Unitary37}
\end{center}
\end{figure}

One strategy of finding solutions to the multi-tensor constraints is to impose that the $A$ and $B$ have some internal structure; that they themselves are formed from a network of smaller tensors. Let us assume that indices on $A$ and $B$ are of some bond dimension $\chi$ such that they can be decomposed as product of four finer indices of dimension $\tilde{\chi} = \chi^{1/4}$. Then we form tensor $A$ from a product of three tensors $Y_{ijkl}$, each of which has four indices of dimension $\tilde{\chi}$, see also Fig. \ref{fig:Parameterization37}(a). The tensors $Y$ are constrained to be simultaneously unitary across both a horizontal and vertical partition of their indices,
\begin{align}
\sum\limits_{kl} {{Y_{ijkl}}Y_{mnkl}^*}  & = {\delta _{im}}{\delta _{jn}}\nonumber \\
\sum\limits_{jl} {{Y_{ijkl}}Y_{mjnl}^*}  & = {\delta _{im}}{\delta _{kn}}.
\end{align}
with $Y^*$ as the complex conjugate of $Y$, see also Fig. \ref{fig:Parameterization37}(b). We call such tensors, which are simultaneously unitary across two partitions, \emph{doubly unitary} tensors. An example of a doubly unitary tensor $Y$, here for a $\mathbb{Z}_2$ symmetric tensor with bond dim $\tilde \chi = 2$, is as follows,
\begin{equation}
{Y_{(ij)(kl)}} = \left[ {\begin{array}{*{20}{c}}
{{c_1}}&0&0&{{s_1}}\\
0&{{s_1}}&{\mathbf{i} {c_1}}&0\\
0&{\mathbf{i}{c_1}}&{{s_1}}&0\\
{{s_1}}&0&0&{{c_1}}
\end{array}} \right],\label{eq:Yparam}
\end{equation}
where $\mathbf{i}$ denotes imaginary unit. The brackets on the indices in ${Y_{(ij)(kl)}}$ are used to denote that the four-index tensor has been reshaped into a matrix according to the grouping shown, and $c_1$ and $s_1$ are shorthand for $\cos(\theta_1)$ and $\sin(\theta_1)$ respectively. Note that, although not specified in the constraints of Fig.\ref{fig:Tiling37}, we have also chosen $Y$ to be reflection symmetric, $Y_{ijkl}=Y_{klij}$, such that the resulting hyper-invariant network will also be symmetric with respect to helicity.

\begin{figure}[!t!b]
\begin{center}
\includegraphics[width=8.6cm]{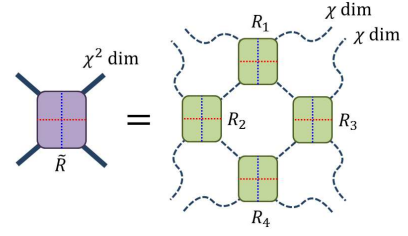}
\caption{If the tensors $\{R_1, R_2, R_3, R_4 \}$ are all doubly unitary, as depicted in Fig. \ref{fig:Parameterization37}(b), then the tensor $\tilde R$ built from combining these tensors (and grouping a pair of $\chi$ dim indices into an index of dimension $\chi^2$) is also doubly unitary. }
\label{fig:DoubleUnitary}
\end{center}
\end{figure}

\begin{figure}[!t!b]
\begin{center}
\includegraphics[width=8.6cm]{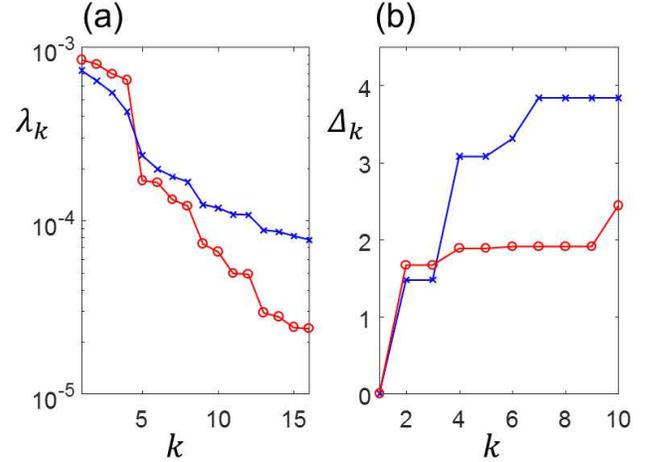}
\caption{(a) Spectra $\lambda_k$ of a 3-site reduced density matrix $\rho$ from the $\{7,3\}$ hyper-invariant network under two different instances of randomly initialised tensors $A$ and $B$, built according to Fig. \ref{fig:Parameterization37}. In both cases, each eigenvalue $\lambda_k$ has a degeneracy of 256 in the full spectrum, which has dimension of $16^3 = 4096$. (b) Scaling dimensions $\Delta_k$, as defined in Eq. \ref{eq:scaling}, of the descending superoperator from Fig. \ref{fig:DescendingSupers}(a) under two different instances of randomly initialised hyper-invariant networks.}
\label{fig:Numeric}
\end{center}
\end{figure}

Similarly, tensors $B$ are formed as the product of a unitary tensor $Q_{ijkl}$ and a doubly unitary tensor $R_{ijkl}$, see Fig. \ref{fig:Parameterization37}(c). We require tensor $R$ to satisfy two sets of unitary constraints, 
\begin{align}
\sum\limits_{kl} {{R_{ijkl}}R_{mnkl}^*}  & = {\delta _{im}}{\delta _{jn}}\nonumber \\
\sum\limits_{jk} {{R_{ijkl}}R_{mjkn}^*}  & = {\delta _{im}}{\delta _{ln}},
\end{align}
see also Fig. \ref{fig:Parameterization37}(e), and to also be reflection symmetric across both axes, $R_{ijkl} = R_{klij}$ and $R_{ijkl} = R_{jilk}$. An example of a tensor $R$, again with $\mathbb{Z}_2$ symmetry and bond dim $\tilde \chi = 2$, that satisfies these constraints is,
\begin{equation}
{R_{(ij)(kl)}} = \left[ {\begin{array}{*{20}{c}}
{{c_2}}&0&0&{\mathbf{i}{s_2}}\\
0&{{c_2}}&{\mathbf{i}{s_2}}&0\\
0&{\mathbf{i}{s_2}}&{{c_2}}&0\\
{\mathbf{i}{s_2}}&0&0&{{c_2}}
\end{array}} \right],\label{eq:Rparam}
\end{equation}
where $c_2$ and $s_2$ are shorthand for $\cos(\theta_2)$ and $\sin(\theta_2)$ respectively. The unitary tensor $Q$ is required to satisfy: 
\begin{equation}
\sum\limits_{kl} {{Q_{ijkl}}Q_{mnkl}^*}  = {\delta _{im}}{\delta _{jn}}
\end{equation}
see also Fig. \ref{fig:Parameterization37}(d), in addition to being reflection symmetric across two axes, $Q_{ijkl} = Q_{klij}$ and $Q_{ijkl} = Q_{jilk}$. An example of a tensor $Q$ that satisfies these constraints, again with $\mathbb{Z}_2$ symmetry and bond dim $\tilde \chi = 2$, is given as,
\begin{equation}
{Q_{(ij)(kl)}} = \left[ {\begin{array}{*{20}{c}}
{{c_3}}&0&0&{{s_3}{e^{\mathbf{i}{\theta _4}}}}\\
0&{{c_5}}&{\mathbf{i}{s_5}}&0\\
0&{\mathbf{i}{s_5}}&{{c_5}}&0\\
{{s_3}{e^{\mathbf{i}{\theta _4}}}}&0&0&{ - {c_3}{e^{2\mathbf{i}{\theta _4}}}}
\end{array}} \right],\label{eq:Qparam}
\end{equation}
where $c_\alpha$ and $s_\alpha$ are shorthand for $\cos(\theta_\alpha)$ and $\sin(\theta_\alpha)$ respectively.

It can then be seen that the $A$ and $B$ tensors constructed in the manner described satisfy the desired multi-tensor constraints (up to an irrelevant multiplicative constant), see also Fig. \ref{fig:Unitary37}. Thus, with Eqs. \ref{eq:Yparam}, \ref{eq:Rparam} and \ref{eq:Qparam}, we have constructed an example of a family of tensors $A$ and $B$, here with $\mathbb{Z}_2$ symmetry and overall bond dim $\chi = 16$, with five free parameters $\{\theta_1,\theta_2,\theta_3,\theta_4,\theta_5 \}$. It is also easily argued that the number of free parameters can be increased as the bond dimension is increased. This follows as a doubly unitary tensor of bond dimension $\chi^2$ can be formed by contracting together any four doubly unitary tensors of bond dimension $\chi$, see Fig. \ref{fig:DoubleUnitary}, although this is not the optimal strategy for generating doubly unitary tensors (in terms of maximising the number of free parameters in the tensors). Thus we have provided an example of tensors $A$ and $B$ that satisfy the first two criteria outlined at the start of this section.

In order to demonstrate that the example tensors satisfy the third criterion, that they describe hyper-invariant tensor networks with non-trivial entanglement properties and correlation functions, we turn to numerics. Firstly, we compute the spectra of 3-site reduced density matrices resulting from two different random choices of the parameters $\{\theta_1,\theta_2,\theta_3,\theta_4,\theta_5 \}$, as plotted in Fig. \ref{fig:Numeric}(a). These results demonstrate that the entanglement properties of these hyper-invariant tensor network are both non-trivial and $\theta$-dependant. Secondly we have examined the spectra of scaling dimensions $\Delta_k$,
\begin{equation}
\Delta_k \equiv -\log_s (\lambda_k) \label{eq:scaling}
\end{equation}
obtained from the eigenvalues $\lambda_k$ of the descending superoperator in Fig. \ref{fig:DescendingSupers}(a) under two different random choices of the parameters $\{\theta_1,\theta_2,\theta_3,\theta_4,\theta_5 \}$, as plotted in Fig. \ref{fig:Numeric}(b). Here the base of the logarithm is the scale factor of the network, $s=(3+\sqrt{5})/2$, as given in Eq. \ref{eq:scalefactor}. As with the scale-invariant MERA, the scaling dimensions $\Delta_i$ with their corresponding scaling operators $\phi_i$ imply that the correlation functions (on certain sites) decay polynomially with distance $l$,  
\begin{equation}
\left\langle {{\phi_i(r)}{\phi_j(r + l)}} \right\rangle  \propto {l^{ - (\Delta_i + \Delta_j) }}.
\end{equation}
Thus the results from Fig. \ref{fig:Numeric}(b) demonstrate that the correlation functions of the hyper-invariant tensor networks are both non-trivial (at all length scales) and $\theta$-dependant. Though not plotted, it was found that the spectra of scaling dimensions from the other superoperators of Fig. \ref{fig:DescendingSupers}, as well as the spectra from various  compositions of superoperators, were also non-trivial (indicating that generic connected correlation functions are non-zero).

\begin{figure}[!t!b]
\begin{center}
\includegraphics[width=8.6cm]{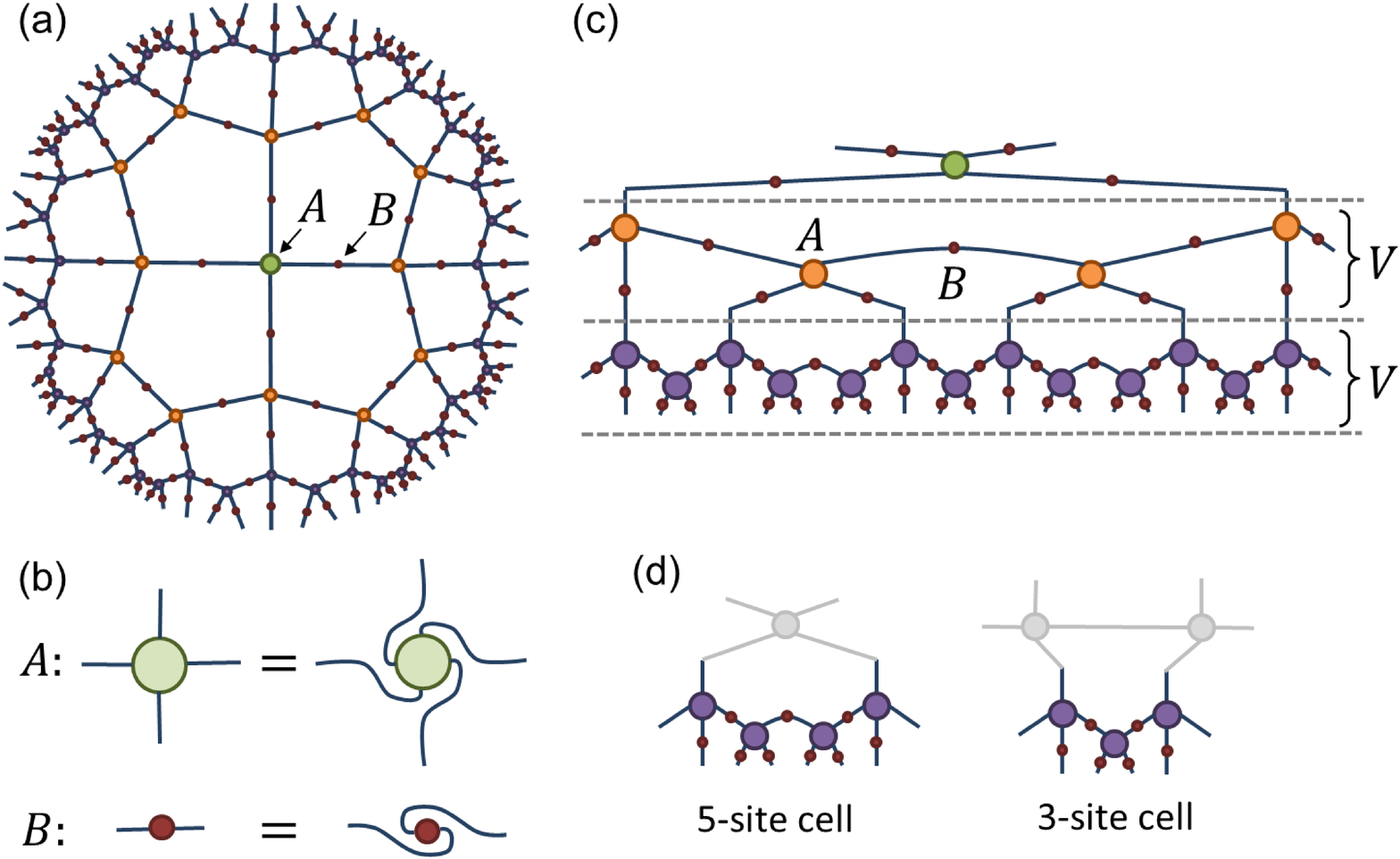}
\caption{(a) A hyper-invariant network constructed according to a $\{5,4\}$ tiling of the hyperbolic disk, where a 4-index tensor $A$ has been placed on each node of the tiling and a matrix $B$ placed on each edge connecting two nodes. (b) Tensors $A$ are constrained to be rotationally invariant and matrices $B$ are constrained to be symmetric. (c) The network can be unwrapped into concentric layers $V$ about any chosen bulk point, where layers are formed from a connected chain of alternating $A$ and $B$ tensors. (d) Each layer is formed from a combination of 5-site and 3-site unit cells.}
\label{fig:Tiling45}
\end{center}
\end{figure}

\begin{figure}[!t!b]
\begin{center}
\includegraphics[width=8.6cm]{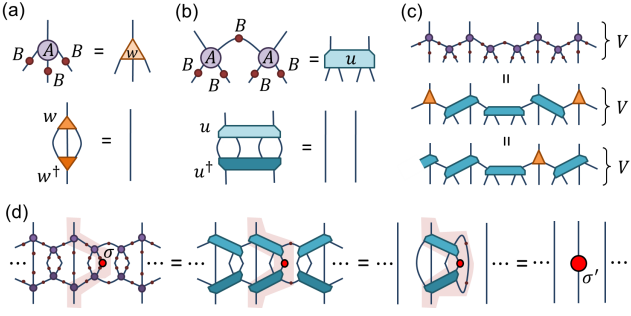}
\caption{Multi-tensor constraints for the $\{5,4\}$ hyper-invariant network. (a) The product of an $A$ tensor with three $B$ matrices is constrained to be a 3-to-1 isometry which, by definition, annihilates to identity with its conjugate $w^\dag$. (b) The product of two $A$ tensors together with five $B$ tensors is constrained to be a 4-to-2 isometry. (c) The tensors in a layer $V$ of the network can be grouped into a product of $w$ and $u$ isometries in many different ways. (d) A one-site local operator $\sigma$ is coarse-grained to a new operator through a layer $V$ of the network, $\sigma' \equiv {V^\dag }\sigma V$. Many tensors in $V$ annihilate with their conjugates in $V^{\dag}$, such that $\sigma'$ here remains a one-site operator. }
\label{fig:Constraint45}
\end{center}
\end{figure}

\section{Section C: Hyper-invariant tensor network based on a \{5,4\} tiling}
Just as there are many forms of MERA for $1D$ lattices, such as the binary and ternary MERA \cite{Evenbly09}, one can similarly construct other forms of hyper-invariant tensor network than the $\{7,3\}$ form considered in the main text. In this section we introduce a hyper-invariant tensor network based on a $\{5,4\}$ tiling of the hyperbolic disk, and demonstrate that it shares the same key properties as the $\{7,3\}$ network.

\begin{figure}[!t!b]
\begin{center}
\includegraphics[width=6cm]{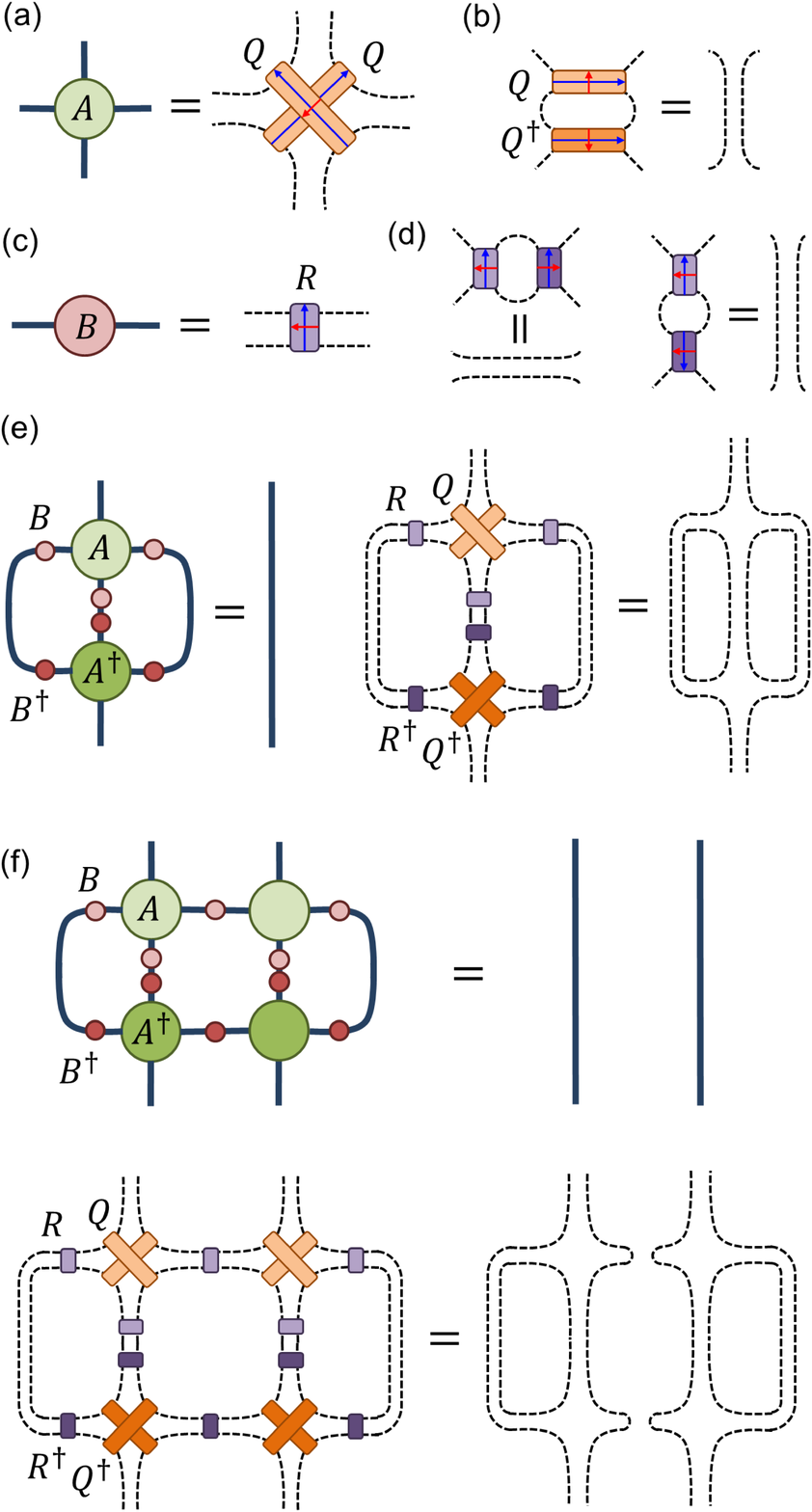}
\caption{An example construction of tensors $A$ and $B$ that satisfy the multi-tensor constraints of Fig. \ref{fig:Constraint45}(a-b). (a) The indices of tensor $A$, assumed o be of dimension $\chi$, are decomposed as a product of two finer indices (dashed) of dimension $\tilde \chi = \sqrt{\chi}$, such that $A$ is then given as the product of two tensors $Q$. (b) Tensors $Q$ are constrained unitary. (c)  Tensors $B$ are formed by grouping pairs of indices together from a 4-index tensor $R$. (d) Tensors $R$ are constrained to be doubly unitary. (e-f) The tensors $A$ and $B$ are seen to satisfy the multi-tensor constraints of the $\{5,4\}$ hyper-invariant network.}
\label{fig:Parameterization45}
\end{center}
\end{figure}

A depiction of the $\{5,4\}$ hyper-invariant tensor network is given in Fig. \ref{fig:Tiling45}(a). The network is formed by placing a 4-index tensor $A$ on each node of a $\{5,4\}$ tiling of the hyperbolic disk, then a matrix $B$ on each edge connecting two $A$ tensors. We constrain the $A$ tensor to be rotationally invariant and the $B$ matrix to be symmetric, see Fig. \ref{fig:Tiling45}(b), such that the network does not have a preferred location or direction in the bulk. The network can be organised in concentric layers $V$ about any chosen bulk location $T$, where each layer, which is a connected chain of alternating $A$ and $B$ tensors, can be resolved as a combination of 3-site and 5-site unit cells, see Fig. \ref{fig:Tiling45}(c-d). In the thermodynamic limit, the ratio $r$ of 3-site to 5-site unit cells and the scale factor $s$, which describes the ratio of indices entering and exiting a layer $V$, are 
\begin{equation}
r = \sqrt{3} \approx 1.732, \; \; \; s = 2+r \approx 3.732.
\end{equation}
As with the $\{7,3\}$ network, we now wish to further constrain the tensors of the $\{5,4\}$ network such that each layer $V$, when interpreted as a transformation between an initial lattice $\L_z$ and a coarser lattice $\L_{z+1}$, preserves locality. Multi-tensor constraints that achieve this are depicted in Fig. \ref{fig:Constraint45}(a-b); here a product of three $B$ tensors together with an $A$ tensor is constrained to form a 3-to-1 isometry $w$, while a product of five $B$ tensors together with two $A$ tensors is constrained to form a 4-to-2 isometry $u$. These constraints allow the tensors in a layer $V$ to be grouped into a product of isometries $w$ and $u$ in many different ways, see Fig. \ref{fig:Constraint45}(c). Given a local operator $\sigma$ on $\L_z$, a coarse-grained operator $\sigma'$ on $\L_{z+1}$ is obtained as,
\begin{equation}
\sigma' \equiv {V^\dag }\sigma V,
\end{equation}
see also Fig. \ref{fig:Constraint45}(d). It can then be seen that if $\sigma$ has a support of $L\le2$ sites on $\L_z$, then the coarse-grained operator $\sigma'$ will have a support of $L\le2$ sites on $\L_{z+1}$. Although we shall not do so here, it can also be argued that the $\{5,4 \}$ hyper-invariant network has the same holographic causality property (and same causal features in general) as the $\{7,3 \}$ network discussed in the main text.

We now provide an example of family of tensors $A$ and $B$ compatible with bulk uniformity that satisfy the multi-tensor constraints of Fig. \ref{fig:Constraint45}(a-b). As with the example from the $\{7,3\}$ network, solutions to the multi-tensor constraints are found by imposing that the $A$ and $B$ tensors have some internal structure. We assume that indices on $A$ and $B$, which are of some bond dimension $\chi$, can be decomposed as product of two finer indices of dimension $\tilde{\chi} = \sqrt{\chi}$. Then tensor $A$ is formed from a product of two tensors $Q_{ijkl}$, each of which has four indices of dimension $\tilde{\chi}$ and is constrained to be unitary, see Fig. \ref{fig:Parameterization45}(a-b), while matrix $B$ is formed from a doubly unitary tensor $R_{ijkl}$, see Fig. \ref{fig:Parameterization45}(c-d). It is then seen in Fig. \ref{fig:Parameterization45}(e-f) that the resulting tensors $A$ and $B$ satisfy the desired multi-tensor constraints. It is worth noting that this example parametrization is much simpler than the example presented for the  $\{7,3 \}$ network, and can be implemented with bond dimension as small as $\chi=4$.

\begin{figure}[!t!b]
\begin{center}
\includegraphics[width=8.6cm]{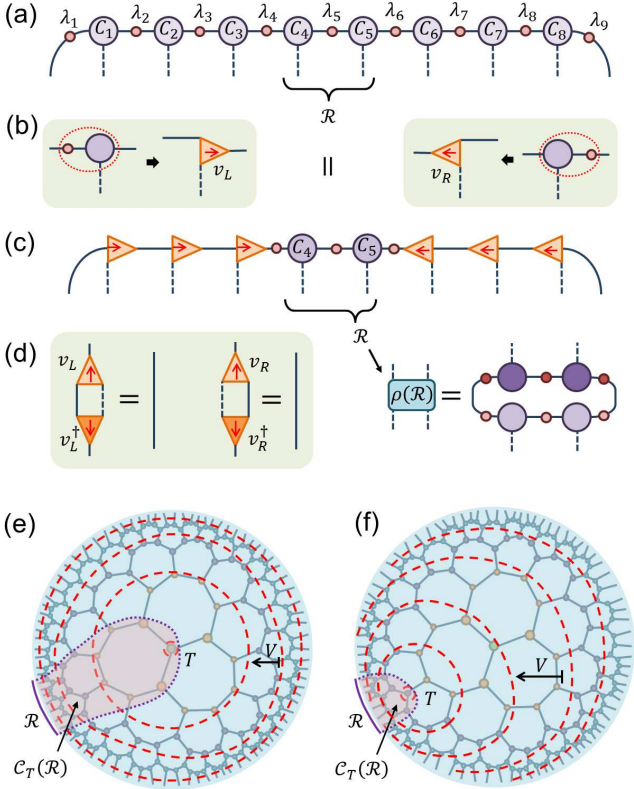}
\caption{(a) A matrix product state in cannonical form, consisting of 3-index tensors $C$ and index weights $\lambda$. (b) Isometries $v_L$ and $v_R$ are formed from contracting $C$ tensors with weights $\lambda$ from the left or right respectively. (c-d) The density matrix $\rho(\R)$ associated to a region $\R$ is constructed from tensors $C$ with local indices in $\R$ (and the weights $\lambda$ on indices connected to these tensors). (e) The hyper-invariant tensor network reproduces the freedom to group the tensors surrounding a chosen center into isometries, but for the hyperbolic disk instead of a $1D$ line, which results in the property of holographic causality.}
\label{fig:Cannonical}
\end{center}
\end{figure}

\section{Section D: Orthogonality centres and cannonical forms}
In this section we further discuss the explanation of holographic causality based on exploiting the freedom to choose the centre $T$ of the layering, which we argue is analogous to the freedom to chose the orthogonality center in matrix product states (MPS).

For MPS there exists a well-defined cannonical form, based on the Schmidt decomposition, with has a number of useful properties. Let us consider a cannonical form MPS on a $1D$ lattice $\L$, composed of 3-index tensors $C$ which sit above each lattice site and diagonal matrices of weights $\lambda$, which sit between the $C$ tensors, see also Fig. \ref{fig:Cannonical}(a). One implication of the cannonical form is that the reduced density matrix $\rho(\R)$ for a continuous region $\R \in \L$ can be constructed from just the tensors $C$ with local indices in $\R$ and the weights $\lambda$ on their indices. This follows as the $C$ tensors outside of $\R$ can be formed into isometries $v_L$ and $v_R$ via left and right multiplication respectively of tensors $C$ by weights $\lambda$, which then annihilate in the contraction of $\rho(\R)$, see also Fig. \ref{fig:Cannonical}(b-d). Notice that the set of tensors that contribute to $\rho(\R)$, which defines the causal cone $\C(\R)$, is also an entanglement wedge $\E(\R)$ of the MPS (provided the region $\R$ is of greater length than $L=1$ sites) as they are contained within the minimal surface $\gamma_\R$. Thus we observe that cannonical form MPS obey an equivalent holographic causality property as discussed for hyper-invariant networks. 

For any tensor network with closed loops there does exist a cannonical form based on the Schmidt decomposition. However the hyper-invariant network can be thought of as generalizing a key aspect of the cannonical form MPS from the $1D$ line to the hyperbolic disk, in that the network may be organised into concentric layers of isometric mappings $V$ about \emph{any} bulk point $T$. The freedom to choose this point $T$, which serves an equivalent role as the center of orthogonality does in an MPS, follows from the multi-tensor constraints imposed on the hyper-invariant network, which may be understood as generalizing isometric constraints of the cannonical form MPS. As commented in the main text, the freedom to choose $T$ results in the holographic causality property of hyper-invariant networks. For a continuous boundary region $\R$ we observe that the size of the apparent causal cone $\C_T(\R)$ is minimized by choosing $T$ at the apex of the minimal surface $\gamma_\R$, for which the apparent causal cone becomes coincident with the true causal cone $\C_T(\R) = \C(\R)$, see Fig. \ref{fig:Cannonical}(e-f). It is also for this choice of center $T$ that $\C_T(\R)$ becomes approximately coincident with the entanglement wedge $\E(\R)$, thus resulting in the holographic causality property.

\begin{figure}[!t!b]
\begin{center}
\includegraphics[width=7.5cm]{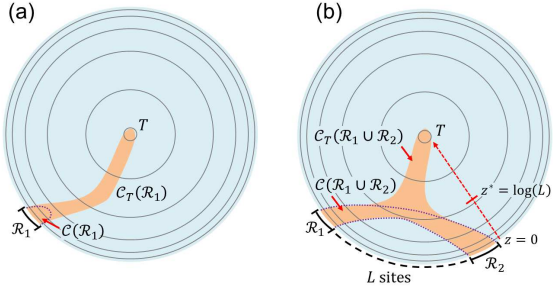}
\caption{(a) The apparent and actual causal cones, $C_T(\R_1)$ and $C(\R_1)$, of a boundary region $\R_1$ in a hyper-invariant network that has been organised into layers about bulk point $T$. (b) The causal cone $C(\R_1 \cup \R_2)$ of disjoint regions $\R_1$ and $\R_2$ is greater than the union of their individual causal cones $C(\R_1)$ and $C(\R_2)$.}
\label{fig:Correlator}
\end{center}
\end{figure}

\section{Section E: Causal cones of disjoint regions}
In this section we briefly discuss the causal cones of disjoint regions from hyper-invariant networks. Let us consider two regions $\R_1$ and $\R_2$, each of $l$ sites, which are separated by distance $L \gg l$ sites on the boundary of a hyper-invariant network. Holographic causality implies that the individual causal cones of $\R_1$ and $\R_2$ are disjoint, $\C(\R_1) \cap \C(\R_2) = \emptyset $, thus one might think that the connected correlation functions between operators supported on these regions should be trivial. This is incorrect, however, as the same manipulations that allow an apparent causal cone $\C_T(\R)$ to be reduced to the true causal $\C(\R)$ for a single boundary region cannot be performed in the case of the disjoint regions. Simply put, here the causal cone of the combined regions is larger than the union of the separate causal cones, i.e. $\C({\R_1} \cup {\R_2}) \ne \C({\R_1}) \cup \C({\R_2})$, in contrast to quantum circuit with a fixed unitary structure where one has $\C({\R_1} \cup {\R_2}) = \C({\R_1}) \cup \C({\R_2})$.

In the present case the apparent causal cone $\C_T({\R_1} \cup {\R_2})$ of the combined regions is connected, and takes the same form as the causal cone of disjoint regions in MERA, where the individual cones $\C_T({\R_1})$ and $\C_T({\R_2})$ fuse at some length scale $z^*\approx \log_s(L)$, see Fig. \ref{fig:Correlator}. In the evaluation of $\C({\R_1} \cup {\R_2})$ only the part of the apparent causal cone $\C_T({\R_1} \cup {\R_2})$ at scales $z>z^*$ cancels, compatible with the existence of non-trivial connected correlations functions. 


\begin{thebibliography}{99}
\bibitem{geodesic}
In the context of tensor networks, minimal surfaces are defined as the surfaces that transect the minimal number of bulk indices. If there is more than one minimal surface for a region $\R$, then $\gamma_{\R}$ is chosen to make the entanglement wedge as large as possible.
\end{thebibliography}
\end{document}